\documentclass[twocolumn,prb]{revtex4-2}
\usepackage{graphicx,psfrag,color}
\def\no{\noindent}
\def\bc{\begin{center}}
\def\ec{\end{center}}

\def\beq{\begin{equation}}
\def\eeq{\end{equation}}

\def\d{\downarrow}
\def\u{\uparrow}

\def\br{{\bf r}}

\def\bk{{\bf k}}

\def\bc{{\bf c}}

\begin{document}

\title{
Anomalous Josephson effect of s-wave pairing states in chiral double layers
}

\author{Klaus Ziegler$^{1}$, Andreas Sinner$^{1}$, and Yurii E. Lozovik$^{2,3}$}
\affiliation{
$^{1}$ Institut f\"ur Physik, Universit\"at Augsburg, D-86135 Augsburg, Germany\\
$^{2}$ Institute of Spectroscopy, Russian Academy of Sciences,142190 Troitsk, Moscow, Russia\\
$^{3}$ Moscow Institute of Electronics and Mathematics, National Research University\\
Higher School of Economics, 101000 Moscow, Russia
}
\date{\today}

\begin{abstract}
We consider s-wave pairing in a double layer of two chiral metals due to interlayer Coulomb interaction,
and study the Josephson effect near a domain wall, where the sign of the order parameter jumps. 
The domain wall creates two evanescent modes at the exceptional zero-energy point, whose superposition 
is associated with currents flowing in different directions in the two layers. Assuming a toroidal geometry, 
the effective Josephson current winds around the domain walls, whose direction is determined by the
phase difference of the complex coefficients of the superimposed zero-energy modes. Thus, the zero-energy
mode is directly linked to a macroscopic current. 
This result can be understood as an interplay of the conventional Josephson current perpendicular 
and the edge current parallel to a domain wall in a double layer of two chiral metals. As a realization 
we suggest the surface of a ring-shaped topological insulator. The duality between electron-electron
and electron-hole double layers indicates that this effect should also be observable in excitonic
double layers. 
\end{abstract}

\maketitle

A Josephson junction is an important tool for probing paired electron states,
since it is sensitive to the phase difference of the pairing order parameter on both sides
of the junction.
The latter generates a current that flows perpendicular to the Josephson junction. 
This effect has been studied in many systems with paired
states, including conventional s-wave superconductors \cite{JOSEPHSON1962251,PhysRevLett.10.230},
unconventional and topological
superconductors \cite{PhysRevB.60.6308,PhysRevB.67.184505,Kwon2004}. 

\begin{figure}[t]
\begin{center}
\includegraphics[width=4cm,height=3cm]{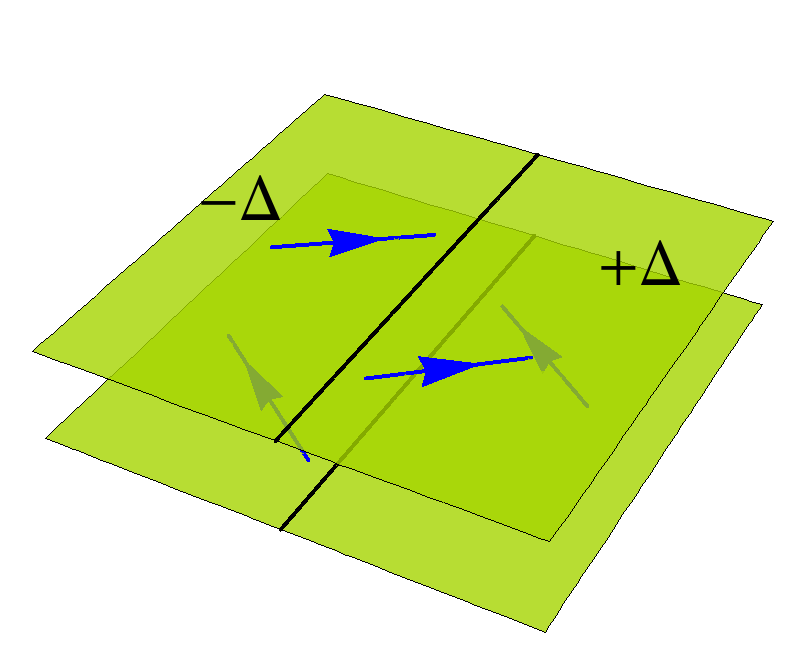}
\caption{
Electronic double layer with domain wall, which is given by a sign jump of the pairing order
parameter. The currents (blue arrows) flow in the same (opposite) direction in the two layers 
parallel (perpendicular) to the domain wall. 
}
\label{fig:domain_wall}
\end{center}
\end{figure}

In this paper we will consider the internal (intralayer) Josephson effect in a two-dimensional 
electronic double layer with interlayer s-wave pairing,
assuming that both layers have a Dirac spectrum consisting of two bands and a spectral node
(see Fig. \ref{fig:domain_wall}).
Such conditions can be realized, for instance, on the surface of a 3D
topological insulator \cite{RevModPhys.83.1057,Burkov_2015}.

The nodal spectrum influences strongly the formation of local currents due to edge modes
along the Josephson junction. They compete with the conventional Josephson current that tends
to cross the junction perpendicularly. As a result we expect an anomalous Josephson effect, where the
effective Josephson current has a component that flows parallel to the junction. The details of these competing 
effects will be studied in this paper. For this purpose the Josephson junction is simplified as 
a domain wall (i.e., it is an infinitesimally narrow Josephson junction). Zero-energy edge modes are created
along the domain wall, causing local currents. Finally, we impose periodic 
boundary conditions to obtain a toroidal geometry with two domain walls.

\begin{figure}[t]
\begin{center}
\includegraphics[width=5cm,height=3.5cm]{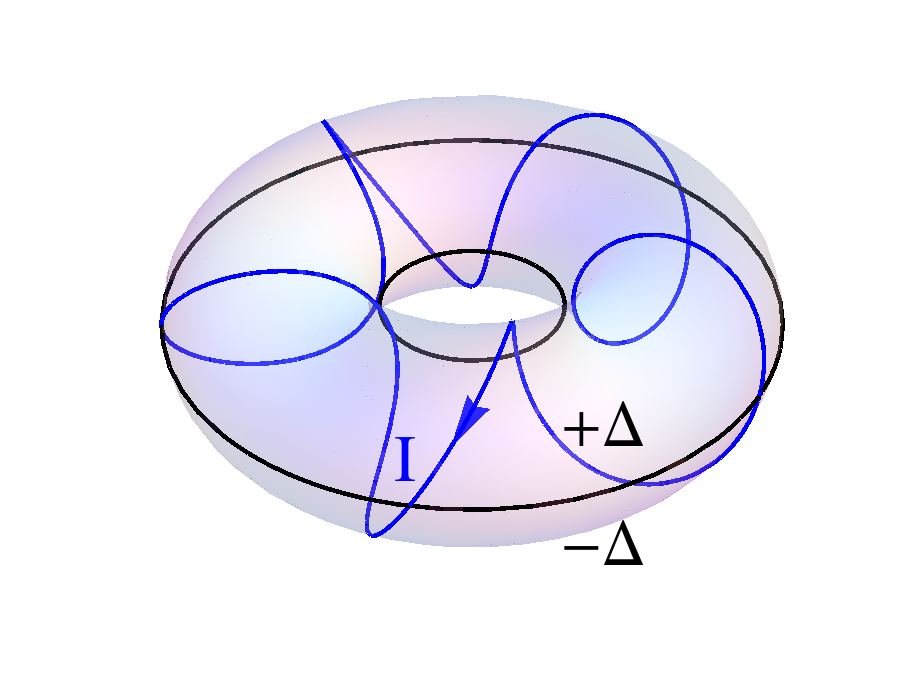}
\caption{
After gluing the double layer of Fig. \ref{fig:domain_wall} to form a torus, the current 
${\bf I}=(j_x+j^s_x,j_y+j^s_y)$ winds along the two domain walls around the torus.
}
\label{fig:torus}
\end{center}
\end{figure}

Our approach is motivated by the fact that
double layers of chiral materials have rich properties due to the combination of electronic interlayer pairing and
quasiparticle edge modes. The interplay of a superconducting Josephson current and edge currents represents an intimate
connection of the Josephson effect and topology, which does not exist in conventional superconductors. Although the edge
modes depend on the sample geometry, the anomalous Josephson effect is robust and determined only by topology, 
which is characterized by the number of edges and domain walls. 
As a special example we will consider in this letter the case of a torus with two domain walls, which has no edges.
Such a geometry can be realized with two metallic layers, separated by a dielectric sheet \cite{Geim2013} 
and connected by a metallic boundary, similar to Fig. \ref{fig:torus}. The internal Josephson current in the double layer
is induced by an external current through inductive coupling. Since the edge currents are directly associated with the
zero-energy quasiparticle states, this gives also a direct access to these states through the external current. 
In particular, it enables us to change the direction of the current relative to the domain wall.
Thus, the sensitivity of the anomalous Josephson effect to an external current provides a method by which one can probe
and control the internal superconducting properties.
This opens a wide field for new experiments on superconducting layered materials.

The theory of the electronic double layer is dual to that of an electron-hole double layer 
due to a duality transformation discussed previously by us \cite{2020PhRvR...2c3085S}.
This relation connects the electronic double layer physics with the excitonic physics.
In particular, the duality suggests that the interlayer Josephson effect should also exist for the 
electron-electron double layer when interlayer hopping is present, as it was
studied before for electron-hole double layers \cite{1976JETP...44..389L,LOZOVIK1997399}. 
In that case the Josephson currents are homogeneous in each layer. 
In the present work this effect will not appear due to the absence of interlayer hopping. 
On the other hand, we expect an anomalous intralayer Josephson effect in electron-hole layers 
when we implement a Josephson junction 
inside the layers, as visualized in Fig. \ref{fig:domain_wall}.

{\it Model:}
We consider an electronic double layer with interlayer Coulomb repulsion
but without interlayer tunneling.
The layers themselves are chiral metals, described by Dirac Hamiltonians with
opposite chirality. Such a system can be defined by the tight-binding Hamiltonian
\[
{\cal H}_{ee}
=\sum_{\br,\br'}\sum_{s=\u,\d}\sum_{\mu,\mu'}H_{\br\br',s,\mu\mu'}c^\dagger_{\br,s,\mu}c^{}_{\br',s,\mu'}
\]
\beq
+\sum_{\br,\br'}\sum_{\mu,\mu'} V_{\br\br'}c^\dagger_{\br,\u,\mu}c^{}_{\br,\u,\mu}
c^\dagger_{\br',\d,\mu'}c^{}_{\br',\d,\mu'}
\ ,
\label{el-el}
\eeq
where $\mu=1,2$ is the band index of the two bands in each layer. $H_{\br\br',\u,\mu\mu'}$
($H_{\br\br',\d,\mu\mu'}$) is the hopping matrix element in the top (bottom) layer, and
$c^\dagger_{\br,\u,\mu}$ creates an electron at site $\br$ in the top layer in the band with 
index $\mu$. An example is the honeycomb lattice,
which is bipartite and consists of two triangular sublattices. In this case $\mu=1,2$ refers
to the two triangular lattices and the coordinate $\br$ refers only to one of the two triangular lattices.
There exists a duality transformation
$c_{\br,\d,\mu}\to d^\dagger_{\br,\d,\mu}$ (i.e., we replace the electrons in the
bottom layer by holes). 
This implies a transformation ${\cal H}_{ee}\to {\cal H}_{eh}$,
where ${\cal H}_{eh}$ is the Hamiltonian of an electron-hole gas  \cite{2020PhRvR...2c3085S}. The latter
interacts via an attractive 
Coulomb interaction, a system that has been studied intensively in terms
of excitons \cite{1976JETP...44..389L,HALPERIN1988711,Eisenstein2004,PhysRevLett.93.266801,
PhysRevB.75.113301,Su2008,PhysRevLett.110.146803,
PhysRevB.86.115436,doi:10.1063/1.4831671,2017NatPh..13..751L,Wang2019}.

{\it Mean-field approximation:}
First, we briefly discuss the  BCS approach for the electron-hole double layer.
For this purpose we introduce the BCS order parameter
\beq
\Delta_{\br\br';\mu\mu'}=V_{\br\br'}\langle c^{}_{\br,\u,\mu}d^{}_{\br',\d,\mu'}\rangle
\ ,
\label{BCS_OP}
\eeq
which describes Cooper pairing of electrons and holes by forming excitons \cite{1976JETP...44..389L}.
Then the interaction term of ${\cal H}_{eh}$ reads in BCS approximation
\beq
\label{bcs_coupling}
-\sum_{\br,\br'}\sum_{\mu,\mu'} V_{\br\br'}c^\dagger_{\br,\u,\mu}c^{}_{\br,\u,\mu}
d^\dagger_{\br',\d,\mu'}d^{}_{\br',\d,\mu'}
\]
\[
\approx
-\sum_{\br,\br'}\sum_{\mu,\mu'}\left( 
d^\dagger_{\br',\d,\mu'}c^\dagger_{\br,\u,\mu}
\Delta_{\br\br';\mu\mu'} + h.c.\right)
\ .
\eeq
The attractive interaction between the electrons in the top layer and the holes in the bottom layer
causes electron-hole interlayer pairing despite the fact that the electronic Coulomb interaction 
$V_{\br\br'}$ is repulsive. 
This leads to the formation of a BCS state because the electron-hole pairs (excitons) can condense.
The interlayer pairing has some similarity with the resonating valence bond idea \cite{ANDERSON1196,fradkin13},
where in the present case the bond consists of the two layers. 

With a uniform order parameter $\Delta$ we get for the 
quasiparticles the Bogoliubov de Gennes Hamiltonian matrix
\beq
\langle {\cal H}_{eh}\rangle\approx H_{BdG}=
\pmatrix{
H_\u & \Delta \cr
\Delta^\dagger & -H^*_\d\cr
}
\ ,
\eeq
where the $2\times2$ matrix structure refers to the top and bottom layer, while $H_{\u,\d}$ and $\Delta$ are 
$2\times 2$ matrices with respect to the band index $\mu$.
This mean-field result can be used to transform back $d^\dagger_{\br,\d,\mu}\to c_{\br,\d,\mu}$,
such that we have again electrons in both layers.
It gives us the effective quasiparticle Hamiltonian matrix
\beq
\langle {\cal H}_{ee}\rangle\approx H_{MF}=\pmatrix{
H_\u & \Delta \cr
\Delta^\dagger & H_\d\cr
}
\ ,
\eeq
which reads for two layers with opposite chiralities
\begin{equation}
H^{}_{MF}
=\pmatrix{
h_1\sigma_1+h_2\sigma_2 & \Delta\sigma_2 \cr
\Delta\sigma_2 & h_1\sigma_1-h_2\sigma_2\cr
}
\ .
\end{equation}
Here we have assumed that the antisymmetric hopping elements $h^{}_1$ and $h^{}_2$
are the same in both layers. 
$\sigma_j$ is the Pauli matrix with respect to the sublattice structure, and $\Delta$ is
a real scalar pairing order parameter. For this Hamiltonian we will discuss the effect of a domain wall.

In Fourier representation $h_{1,2}$ and $\Delta$ can depend
on the 2D wave vector $\bk=(k_x,k_y)$. 
This Hamiltonian belongs to the symmetry class DIII according to Ref.~\cite{PhysRevB.78.195125},
and it has two degenerate bands with first Chern numbers $\pm 1$.
Its gapped quasiparticle dispersion reads $E_\bk=\pm \sqrt{h_1^2+h_2^2+\Delta^2}$.
The latter agrees with the dispersion of the Hamiltonian with identical layers \cite{2020PhRvR...2c3085S}
\begin{equation}
H_{MF}' 
=\pmatrix{
h_1\sigma_1+h_2\sigma_2 & \Delta\sigma_3 \cr
\Delta\sigma_3 & h_1\sigma_1+h_2\sigma_2\cr
}
\ .
\end{equation}
Thus, the difference in terms of chirality can only be seen in the eigenfunctions but not in their spectra. 
It should be noted that in this case the pairing takes place between the same sub-bands, whereas
in the case with opposite chirality the pairing occurs between different sub-bands.

The dispersion $E_\bk=\pm \sqrt{h_1^2+h_2^2+\Delta^2}$ can vanish when
$h_1$ and/or $h_2$ are imaginary. This is the
case for evanescent modes. The phenomenon is known from conventional Josephson junctions,
where evanescent modes exist inside the gap.

{\it Zero-energy modes at a domain wall:}
An inhomogeneous order parameter $\Delta$ breaks translational invariance
and divides the layers in different regions. 
For a Josephson junction \cite{JOSEPHSON1962251} we typically choose 
a region around $x=0$, where the order parameter vanishes.
For our purpose we can consider the simple case \cite{2000PhRvB..6110267R,fradkin13}
that ${\rm sgn}(\Delta)$ jumps at the domain wall along the $y$ direction at $x=0$,
as visualized in Fig. \ref{fig:domain_wall}. 
Such a discontinuous phase change suppresses the Andreev states except for the zero-energy modes.
It should be kept in mind here that the domain wall is created by a potential, and
the corresponding change of the order parameter should be obtained via the BCS-like equation.
In practice, this requires some tedious calculations and we simply focus here
on the domain wall in terms of $\Delta(x)$, following the recipe proposed in 
Ref.~[\onlinecite{2000PhRvB..6110267R}]. Next, we analyze the effect of the domain
wall. The system is translational invariant in $y$ direction, such that we can use Fourier components
with respect to $k_y$. The domain wall breaks the translational invariance in $x$ direction. Considering only the
low energy MF Hamiltonian we can write $h_1\sim i\partial_x$, $h_2\sim k_y$ and
\begin{equation}
H^{}_{MF} 
=\pmatrix{
i\partial_x\sigma_1+h_2\sigma_2 & \Delta(x)\sigma_2 \cr
\Delta(x)\sigma_2 & i\partial_x\sigma_1-h_2\sigma_2\cr
}
\ .
\end{equation}
For the zero-energy mode we can make the ansatz $\Psi_{k_2}(x)=\psi_{k_2}e^{-bx}$,
where $b$ depends on the sign of $x$. Then we get the eigenmode equation
\beq
\pmatrix{
-ib\sigma_1+h_2\sigma_2 & \Delta(x)\sigma_2 \cr
\Delta(x)\sigma_2 & -ib\sigma_1-h_2\sigma_2\cr
}\Psi_{k_2}(x)
=0
\label{H-matrix}
\ .
\eeq
Solving this equation for $x<0$ and for $x>0$ and
using the matching condition at $x=0$, the evanescent solutions require 
$b={\rm sgn}(x)\sqrt{h_2^2+\Delta^2}$ and $h_2=0$, such that $b={\rm sgn}(x)|\Delta|$ and
the zero-energy modes read
\beq
\Psi_1=
\frac{1}{{\cal N}}\pmatrix{
1 \cr
0 \cr
1 \cr
0 \cr
}e^{-|\Delta||x|}
\ ,\ \ 
\Psi_2=\frac{1}{{\cal N}}\pmatrix{
0 \cr
1 \cr
0 \cr
-1 \cr
}e^{-|\Delta||x|}
\label{zero_modes02}
\eeq
with the normalization ${\cal N}=\sqrt{2/|\Delta|}$. $\Psi_1$ has the same wavefunction on the
top and the bottom layer, whereas $\Psi_2$ has an opposite sign on the two layers.

For a more general phase change $\exp(-i\theta/2)\to\exp(i\theta/2)$ at $x=0$, the matching condition reads
\beq
b_-=b_+e^{-i\theta}=b_+\left(\cos\theta-i\sin\theta\right)
\ ,
\label{match1}
\eeq
where $b_+$ ($b_-$) refer to the right (left) side with respect to the domain wall.
To get exponentially decaying functions $\exp(-b_\pm x)$ on both sides of the domain wall, the sign of $Re(b_-)$ 
($Re(b_+$)) must be negative (positive). This requires $\cos\theta<0$ due to Eq. (\ref{match1}); i.e., for
 $\pi/2<\theta<3\pi/2$. The decay length of the bound state is $-1/|\Delta|\cos\theta$ in units of the lattice spacing.
For $\cos\theta\ge0$ there is no bound state at the domain wall. 


{\it Symmetries and degenerate zero-energy modes:}
The characteristic polynomial of the MF Hamiltonian has four degenerate zero solutions for $h_2=0$ and 
$b(x)={\rm sgn}(x)|\Delta|$. 
We consider the block-diagonal matrices $S_j=diag(\sigma_j,\sigma_j)$ and $T_j=diag(\sigma_j,-\sigma_j)$
for $j=1,2,3$.
First, for $h_2=0$ the MF Hamiltonian is invariant under the transformation $H_{MF}\to T_1H_{MF}T_1$, 
which creates a new zero mode $\Psi_2$ from $\Psi_1$ as $\Psi_2=T_1\Psi_1$.
Moreover, $T_2$ as a sublattice transformation is also a particle-hole transformation, since $T_2H_{MF}T_2=-H_{MF}$
and, therefore, $T_2\Psi_E=\Psi_{-E}$.
Thus, $T_1$ and $T_2$ create from $\Psi_1$ three more zero modes. In particular, the fact that
the transformation matrices obey the following rules
\beq
S_j^2=T_j^2={\bf 1}
\ ,\ \ 
T_1T_2=S_1S_2=iS_3
\eeq
and 
$S_3\Psi_1=\Psi_1$, $T_2\Psi_1=i\Psi_2$ and $T_2T_1\Psi_1=S_2S_1\Psi_1=i\Psi_1$ reflects
that the zero energy is an exceptional point with a two-dimensional eigenspace 
\cite{Kato:101545}. In the context of line defects in the BdG Hamiltonian the appearance of exceptional
points has been discussed recently \cite{Mandal_2015}.
It should be noted that the zero eigenmodes of $H_{MF}$ in Eq. (\ref{zero_modes02}) are real.
But any superposition
of the two zero modes $\Phi=a_1\Psi_1+a_2\Psi_2$ with complex coefficients
$a_j=|a_j|e^{i\varphi_j}$ and normalization $|a_1|^2+|a_2|^2=1$ is also a zero mode.
Thus, the zero eigenmodes are complex in general but can also be chosen as real.
Since both zero eigenmodes decay exponentially with $|x|$, 
we must employ an additional condition that creates a unique solution for the physical system. The $y$ direction
does not select valid solutions because we need $h_2=0$ for the matching condition; i.e., the solution must be
constant in that direction, which can only be satisfied by periodic boundary conditions.
A possible solution is to impose a condition on the physical properties, for instance, by fixing 
the current density. This will be discussed subsequently.

{\it Currents:}
The current densities induced by the zero modes are expressed separately
for the top and for the bottom layer, where we have the wavefunctions
\beq
\Phi_\u=\pmatrix{
a_1 \cr
a_2 \cr
}\frac{e^{-|\Delta||x|}}{{\cal N}}
\ ,\ \ 
\Phi_\d=\pmatrix{
a_1 \cr
-a_2 \cr
}\frac{e^{-|\Delta||x|}}{{\cal N}}
\ .
\label{layer_phi}
\eeq
Using the quasiparticle continuity equation \cite{PhysRevB.25.4515}, we get for the top layer
\beq
\partial_t\Phi_\u\cdot\Phi_\u
+\partial_xj_{x\u}
=i\Delta(x)\Phi_\u\cdot\sigma_2\Phi_\d
\label{continuityu}
\eeq
and for the bottom layer
\beq
\partial_t\Phi_\d\cdot\Phi_\d
+\partial_xj_{x\d}
=i\Delta(x)\Phi_\d\cdot\sigma_2\Phi_\u 
\ ,
\label{continuityd}
\eeq
where the scalar product contains an implicit complex conjugation: 
$\Phi_\u\cdot\Phi_\u \equiv|\Phi_{\u1}|^2+|\Phi_{\u2}|^2$.
The current densities are related to the current operator $\frac{e}{i\hbar}[H_{MF},r_\mu]$,
projected onto the top or bottom layer, respectively:
\beq
j_{x\u}=-j_{x\d}
=|\Delta||a_1a_2|\cos(\varphi_2-\varphi_1)e^{-2|\Delta||x|}
\ .
\label{curr_x}
\eeq
The terms on the right-hand side of Eqs. (\ref{continuityu}),  (\ref{continuityd})
represent the source/drain provided by the pairing condensate~\cite{PhysRevB.25.4515},
which can be identified with the supercurrents $j^s_{x\sigma}$ through the relation
\beq
\partial_{x}j^s_{x\u}=-i\Delta(x)\Phi_\d\cdot\sigma_2\Phi_\u
\eeq
and accordingly for $j^s_{x\d}$.
An $x$--integration of this equation from the domain wall to some position $x$ then provides
\beq
j^s_{x\u,\d}(x)\sim \pm|\Delta||a_1a_2|\cos(\varphi_2-\varphi_1)(1-e^{-2|\Delta||x|})
\ .
\eeq
Since the wavefunction is constant with respect to the $y$ component, the corresponding current densities
\beq
j_{y\u}=j_{y\d}=
|\Delta||a_1a_2|\sin(\varphi_2-\varphi_1)e^{-2|\Delta||x|}
\label{curr_y}
\eeq
do not appear in the continuity equations. The properties $j_{y\u}=j_{y\d}$ ($j_{x\u}=-j_{x\d}$)
reflect the fact that the currents in the two layers are (anti-) correlated (cf. Fig. \ref{fig:domain_wall}).
This effect should be experimentally observable, since the interlayer 
current-current correlation is associated with the drag effect \cite{2020PhRvR...2c3085S,1996PhRvL..76.2786V}.

A non-vanishing current requires that both zero modes contribute (i.e. $a_1, a_2\ne 0$). This implies
that according to Eq. (\ref{layer_phi}) the eigenvectors $\Phi_\u$ and $\Phi_\d$ are linearly independent.
The currents of the two layers in $y$ direction are not balanced in contrast to the currents in $x$ direction; 
i.e., they do not cancel each other in the two layers. Thus, there is a net current along the $y$ direction in 
the double layer (cf. Fig. \ref{fig:domain_wall}). 
Since the layers are charge separated, there is no charge current between them. Therefore, the 
currents must be conserved in each layer. 

The picture in Fig. \ref{fig:domain_wall} is incomplete though because the edges of the layers have
not been included. 
But since the wave functions decay exponentially away from the domain wall and the order parameter is the same in 
both layers, we have effectively periodic boundary conditions in $x$ and $y$ direction, resulting in the toroidal geometry
of Fig. \ref{fig:torus}.

After preparing the wavefunction $\Phi=(\Phi_\u,\Phi_\d)$ we create a current density $(j_{x,\u},j_{y,\u})$ in 
the top and $(-j_{x,\u},j_{y,\u})$ 
in the bottom layer. By changing the wavefunction $\Phi$ through the coefficients $a_j$ these current densities 
also change according to the above relations. This change can be achieved experimentally with an external 
current source that couples inductively to the double layer. By choosing the direction of the current 
(i.e., the angle $\varphi_2-\varphi_1$), we excite the corresponding wave function $\Phi$.


{\it Discussion:}
The existence of edge modes, a consequence of the chiral metallic layers,
affects the Josephson currents near the domain wall. A measure of the interplay between
the conventional Josephson current and the edge current is the direction of the quasiparticle current
with respect to the domain walls in Fig. \ref{fig:torus}.
Such edge modes appear on all edges including the sample boundaries. In Fig. 
\ref{fig:domain_wall} the latter have not been depicted because
an infinite 2D sample was assumed. On the other hand, it is well known
that a consistent description of edge modes requires a compact manifold \cite{fradkin13}.
Fig. \ref{fig:torus} presents a compact version of the double layer as a single layer 
on a torus (i.e., for periodic boundary conditions), which has two domain walls. Such 
a geometry can be realized as the surface of a ring-shaped topological insulator.

The existence of two degenerate zero modes requires an additional physical constraint to
lift the degeneracy and to obtain a unique solution. In our case this was achieved by considering the current
in the sample. Depending on that current we get a specific linear combination of the two zero 
modes. From a physical perspective this means that we induce a current density in the system by coupling it
to an external current, which excites the corresponding quasiparticle state.
%
The exponential decay of quasiparticle modes and their corresponding currents away from the domain walls 
on the scale $1/|\Delta|$
implies the existence of a supercurrent due to charge conservation \cite{PhysRevB.25.4515}.
In other words, the quasiparticle currents near the domain wall are part of a stationary current
inside the entire torus, winding around the domain walls (cf. Fig. \ref{fig:torus}). This current,
which is proportional to $|\Delta|$, could
be measured through inductance, e.g., by using a coil. Very accurate measurements
of the current can be performed with a SQUID \cite{PhysRevLett.94.166802}. A more direct probe of 
the zero modes would be possible with an electronic Mach-Zehnder interferometer 
\cite{Ji2003,PhysRevLett.102.216404}.

Our calculation was performed for the special example of a jump of the order parameter phase $\pm\Delta$.
Similar calculations can be done for other shapes of $\Delta=|\Delta|e^{i\alpha(x)}$, provided that the order
parameter phase represents a kink with a global phase change from one boundary to the other
in $x$ direction.
Although the anomalous Josephson effect will not be changed qualitatively, the decay of the
wavefunctions and of the current densities is increased to $-1/|\Delta|\cos\theta$ for a phase change $\theta$.
Moreover, a  broader Josephson junction might have different modes (cf. 
discussion in Ref. [\onlinecite{2018RSPTA.37680140S}]).

In conclusion, we found an anomalous Josephson current caused by the superposition of two zero energy
modes in the vicinity of the domain wall. The direction of this current is determined by the
coefficients of the superposition. Conversely, an external current in a specific direction can induce a 
certain superposition of the zero modes on the surface of a ring-shaped topological insulator.

\vskip0.5cm

\no
{\bf Acknowledgments:}

\no
This research was supported by a grant of the Julian Schwinger Foundation for Physics Research.
Yu.E.L. was supported through the grants RFBR 20-02-00410 and 20-52-00035.

\bibliography{bdg_app}

\end{document}